\documentclass[onecolumn,secnumarabic,amsmath,amssymb,balancelastpage,nofootinbib]{article}

\usepackage[e]{esvect}
\usepackage{color}         % produces boxes or entire pages with colored backgrounds
\usepackage{graphics}      % standard graphics specifications
\usepackage{graphicx}      % alternative graphics specifications
\usepackage{epsf}          % old package handles encapsulated post script issues
\usepackage{bm}            % special 'bold-math' package

\usepackage{cite}                                     
\usepackage{amssymb}
\usepackage{amsmath, esint}
\usepackage{mathrsfs}
\usepackage{framed}
\usepackage{bigints} % for big integral symbol
\usepackage{enumitem}
\usepackage{sidecap} % for side captioning figures
\usepackage{pifont}
\usepackage[capbesideposition={left,center},facing=yes,capbesidewidth=8cm,capbesidesep=quad]{floatrow} % for side captioning
\usepackage{setspace} % for double or single spacing

\usepackage[none]{hyphenat} % removes hyphenation

\usepackage[colorlinks=true]{hyperref}  % this package should be added after all others

\usepackage{feynmf}

\definecolor{darkred}{rgb}{0.6,0,0}
\definecolor{darkgreen}{rgb}{0,0.5,0}
\definecolor{darkblue}{rgb}{0,0,0.6}
\hypersetup{ colorlinks,
linkcolor=darkblue,
filecolor=darkgreen,
urlcolor=darkgreen,
citecolor=darkred }

\newcommand{\del}[0]{\ensuremath{\vec{\nabla}}}

\begin{document}

\sloppy % stops it from having words stick out to the right because I forbade hyphenation

\bibliographystyle{nar}

%The block below compresses the space between items in the bibliography
\newlength{\bibitemsep}\setlength{\bibitemsep}{.2\baselineskip plus .05\baselineskip minus .05\baselineskip}
\newlength{\bibparskip}\setlength{\bibparskip}{0pt}
\let\oldthebibliography\thebibliography
\renewcommand\thebibliography[1]{%
  \oldthebibliography{#1}%
  \setlength{\parskip}{\bibitemsep}%
  \setlength{\itemsep}{\bibparskip}%
}

%\numberwithin{equation}{section} %sets equation numbers <chapter>.<section>.<index>
%\numberwithin{figure}{section}

\title{\vspace*{-35 pt}\huge{How Anomalous is the Electron's\\Magnetic Moment?}}
\author{Charles T. Sebens\\Division of the Humanities and Social Sciences\\California Institute of Technology}
\date{arXiv v1 - April 29, 2025\\Forthcoming in \emph{Foundations of Physics}}

\maketitle
\vspace*{-20 pt}
\begin{abstract}
The electron's spin magnetic moment is ordinarily described as anomalous in comparison to what one would expect from the Dirac equation.  But, what exactly should one expect from the Dirac equation?  The standard answer would be the Bohr magneton, which is a simple estimate of the electron's spin magnetic moment that can be derived from the Dirac equation either by taking the non-relativistic limit to arrive at the Pauli equation or by examining the Gordon decomposition of the electron's current density.  However, these derivations ignore two effects that are central to quantum field theoretic calculations of the electron's magnetic moment: self-interaction and mass renormalization.  Those two effects can and should be incorporated when analyzing the Dirac equation, to better isolate the distinctive improvements of quantum field theory.  Either of the two aforementioned derivations can be modified accordingly.  Doing so yields a magnetic moment that depends on the electron's state (even among $z$-spin up states).  This poses a puzzle for future research: How does the move to quantum field theory take you from a state-dependent magnetic moment to a fixed magnetic moment?
\end{abstract}

\newpage
\tableofcontents
\newpage

\section{Introduction}

Making certain assumptions, one can derive a simple expression for the (spin) magnetic moment of the electron from the Dirac equation: the Bohr magneton.  This serves as a decent approximation to the true magnetic moment of the electron.\footnote{This paper is focused on the spin magnetic moment of the electron, not its orbital or total magnetic moment.  I sometimes drop the word ``spin" and use ``magnetic moment'' as shorthand for spin magnetic moment.}  In reality, the electron's magnetic moment is not exactly equal to the Bohr magneton---it is slightly stronger.  The difference between the electron's actual magnetic moment and the Bohr magneton is called its ``anomalous magnetic moment.''  Ordinarily, one goes to quantum field theory to calculate the electron's anomalous magnetic moment, and there it is possible to do so to astounding accuracy through the arduous evaluation of Feynman diagrams.\footnote{The anomalous magnetic moment can be calculated to high precision within quantum electrodynamics and to higher precision within the full standard model \cite{gabrielse2013, koberinski2020}.}  This has frequently been touted as one of the most accurate predictions in physics.  For example, Gabrielse \cite{gabrielse2013}---an experimentalist who has conducted precision measurements of the electron's anomalous magnetic moment---calls a particular measurement of the electron's magnetic moment  ``the most precise confrontation of any theory and experiment'' in an article boldly titled ``The Standard Model’s Greatest Triumph.''  This situation prompts a natural desire to understand why it is that in quantum field theory the electron gains a small anomalous magnetic moment as compared to the value predicted by the Dirac equation (the Bohr magneton).

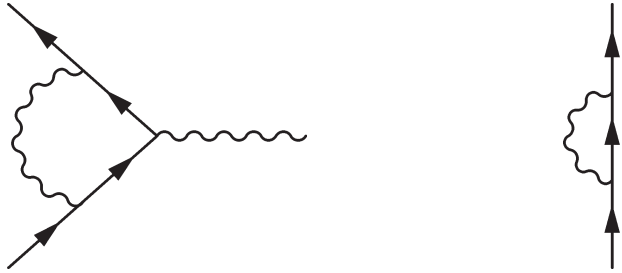
\begin{figure}[htb]
\center{
\begin{fmffile}{vertex}
\begin{fmfgraph*}(125,100)
\fmfleft{a,e}
\fmfright{f}
\fmf{fermion,straight}{a,b,c,d,e}
\fmf{photon,left,tension=0}{b,d}
\fmf{photon}{c,f}
\end{fmfgraph*}
\begin{fmfgraph*}(125,100)
\fmfleft{}
\fmfright{a,d}
\fmf{fermion,straight}{a,b,c,d}
\fmf{photon,left,tension=0}{b,c}
\end{fmfgraph*}
\end{fmffile}
}
\caption{Here are two Feynman diagrams that are used to calculate the anomalous magnetic moment of the electron in quantum electrodynamics, with time going from bottom to top.  Both diagrams show an electron interacting with a photon that it earlier emitted.  The second diagram is associated with mass renormalization in general \cite[sec.\ 9.6]{ryder1996}; \cite[ch.\ 17 and 18]{schwartz}.  The first diagram is used to calculate the anomalous magnetic moment in particular \cite[sec.\ 6.3]{peskinschroeder}; \cite[ch.\ 17]{schwartz}; \cite{harlander2024}.}
\label{feynmanfigure}
\end{figure}

The calculations of the electron's anomalous magnetic moment that are performed within quantum field theory\footnote{For textbook treatments of the first-order correction to the electron's magnetic moment in quantum electrodynamics, see \cite[pg.\ 184--196]{peskinschroeder}; \cite[pg.\  343--345]{ryder1996}; \cite[ch.\ 17]{schwartz}.  Schwinger \cite{schwinger1948, schwinger1949} was the first to present the correction, though it was independently arrived at by Feynman and Tomonaga \cite[pg.\ 320--321]{schwartz}.  Luttinger \cite{luttinger1948} gave an interesting early alternative derivation within quantum electrodynamics that is more similar to the analysis presented here (in section \ref{BDsection}).} have two important features that I would like to highlight here:  First, they involve the study of self-interaction.  This is apparent when you evaluate Feynman diagrams that depict an electron emitting a photon that it itself later absorbs (figure \ref{feynmanfigure}).  Second, these calculations make use of mass renormalization---distinguishing the bare mass of the electron from the mass that it has when you include a contribution associated with the electromagnetic field that surrounds it.

Normally, quantum field theory is viewed as explaining why the electron's actual magnetic moment is a bit stronger than the value predicted by the Dirac equation, where that is understood to be the Bohr magneton.  But, here we are unfairly comparing quantum field theory with self-interaction and mass renormalization to the Dirac equation without either self-interaction or mass renormalization.  To understand what quantum field theory itself contributes to our understanding of the anomalous magnetic moment, we must ask whether these effects can be incorporated into our study of the Dirac equation.  We will see that they can, and that when you do so you get a state-dependent (size-dependent) magnetic moment for the electron (even when you restrict your attention to, say, purely $z$-spin up states).  Quantum field theory still provides a critical improvement over the Dirac equation, but not because it explains a small anomaly in the value of the electron's magnetic moment.  What quantum field theory explains, somehow, is why the electron has a fixed magnetic moment at all, and what that magnetic moment is.\footnote{Does the electron truly have a fixed magnetic moment in quantum field theory?  For our purposes here, the key point is that---despite any state-dependence that might remain---quantum field theory somehow manages to assign a precise magnetic moment to the electron.  The analysis of the Dirac equation presented here does not yield any precise magnetic moment.  The magnetic moment is thoroughly state-dependent.}  How does it explain that?  This is a puzzle that I pose here and leave for future research to solve.

The purpose of this article is to clarify the contribution that quantum field theory makes to our understanding of the electron's magnetic moment by seeing how far we can go within a precursor to quantum field theory.  There are two distinct ways to interpret the Dirac equation as part of a precursor to quantum field theory (as part of a simpler theory that one can modify to arrive at quantum electrodynamics).\footnote{See \cite[ch.\ 11]{durr2020}; \cite{fields}; \cite[sec.\ 6.5]{tumulka2022}.}  First, you might interpret the $\psi$ that appears in the Dirac equation as a quantum wave function that is interacting with a classical electromagnetic field.  Second, you might interpret $\psi$ as a classical charged field interacting with a classical electromagnetic field.  Both interpretive options will be discussed, though the focus will be on equations that are common ground between the two alternatives.

Although we will not be reviewing the way that the anomalous magnetic moment is ordinarily calculated within quantum field theory here, there are reasons to be dissatisfied by the standard treatment.  The comparison with a precursor to quantum field theory (pursued here) could potentially point towards a better method.  Here are some reasons for dissatisfaction:  First, one normally proceeds indirectly by asking how the energy of an electron changes when it is placed in an external magnetic field, instead of simply examining a lone electron itself and determining its magnetic moment.  Second, the Feynman diagrams used to calculate the anomalous magnetic moment depict processes that unfold over time, whereas the electron's magnetic moment should be a feature possessed by the electron at a moment.  Third, Feynman diagrams treat the electron as a point charge whereas there are (arguably) good reasons to believe that the electron's charge is in fact spread out and that it is the rotation of this charge that gives rise to the electron's magnetic moment \cite{ohanian1986, howelectronsspin, smallelectronstates, spinmeasurement, fields}.  Fourth, the calculation from Feynman diagrams does not yield a clear physical picture as to the origin of the anomalous magnetic moment.\footnote{Here is Feynman himself:
\begin{quote}
``It seems that very little physical intuition has yet been developed in this subject [quantum electrodynamics]. \dots\ To make my view clearer, consider, for example, the anomalous electron moment \dots\ We have no physical picture by which we can easily see that the correction is roughly $\alpha/2\pi$, in fact, we do not even know why the sign is positive (other than by computing it). \dots\ We have been computing terms like a blind man exploring a new room, but soon we must develop some concept of this room as a whole, and to have some general idea of what is contained in it.''
Feynman \cite[pg.\ 75--76]{feynman1962} (quoted in \cite[pg.\ 583]{macgregor1989})
\end{quote}
Although this quote is from long ago, our physical intuition has not improved greatly in the intervening decades.
}

The plan for the paper is as follows:  In section \ref{M1section}, we will review the standard method for deriving the electron's spin magnetic moment from the Dirac equation, taking the non-relativistic limit in the presence of an external magnetic field and arriving at the Pauli equation.  This yields the Bohr magneton as the electron's spin magnetic moment, missing any anomalous magnetic moment.  In section \ref{M2section}, we will see an alternative derivation of the Bohr magneton as the spin magnetic moment for the electron from an analysis of the flow of charge around a central axis (as in classical electromagnetism).  This derivation uses the Gordon decomposition of the electron current density and again takes the non-relativistic limit, though for this derivation there is no need to introduce an external magnetic field.  Next, that derivation will be modified to account for self-interaction (section \ref{SICDsection}) and mass renormalization (section \ref{MRsection}).  In section \ref{RM1section}, we will return to the earlier derivation from section \ref{M1section} and see that self-interaction and mass renormalization can similarly be incorporated there.  These analyses will be compared to related approaches by other authors in section \ref{comparisonsection}, focusing on the pioneering work of Barut and his collaborators (who also consider corrections from self-interaction and mass renormalization in a similar context, but ultimately arrive at different results).

\section{Deriving the Dirac Magnetic Moment}\label{Dsection}

The anomalous magnetic moment of the electron is normally presented in quantum field theory textbooks as being anomalous relative to what one would predict from the Dirac equation: the Bohr magneton, a.k.a, the Dirac magnetic moment,\footnote{See, e.g., \cite[pg.\ 188]{peskinschroeder}; \cite[pg.\ 43, 344]{ryder1996}.}
\begin{equation}
\mu_B=\frac{e\hbar}{2 m c}
\ ,
\label{bohrmagneton}
\end{equation}
where $m$ is the mass of the electron and $-e$ is the charge.  Here and throughout, Gaussian CGS units are used.  To first-order, the anomalous magnetic moment calculated from quantum field theory changes the spin magnetic moment to
\begin{equation}
\mu= \frac{e\hbar}{2mc}\left(1 + \frac{1}{2 \pi} \frac{e^2}{\hbar c}\right)
\ ,
\label{anomalousmagneticmoment}
\end{equation}
making it a bit stronger than the Dirac magnetic moment.  In \eqref{anomalousmagneticmoment}, $\frac{e^2}{\hbar c}$ is the fine-structure constant, $\alpha \approx 1/137$.

This section presents two ways of deriving the Dirac magnetic moment from the Dirac equation.  First, there is the standard way of doing so by taking the non-relativistic limit of the Dirac equation, arriving at the Pauli equation, and examining the coupling to an external magnetic field.\footnote{This kind of standard derivation is presented in, e.g., \cite[sec.\ 1.4]{bjorkendrell}; \cite[sec.\ 33]{lifshitzRQM}; \cite[pg.\ 567--569]{shankar1994} \cite[sec.\ 2.6]{ryder1996}; \cite[pg.\ 124--126]{greiner2000}; \cite[pg.\ 500--501]{sakurai2011}.  The derivation in Dirac's own textbook is similar \cite[pg.\ 265]{dirac}.}  Second, there is an alternative method analyzing the flow of charge.

\subsection{Method 1: The Non-Relativistic Limit of the Dirac Equation}\label{M1section}

In classical electromagnetism, the potential energy of a point magnetic dipole $\vec{m}$ in an external magnetic field $\vec{B}$ is
\begin{equation}
U=-\vec{m}\cdot\vec{B}
\ ,
\label{EMpotentialenergy}
\end{equation}
an expression that is minimized when the magnetic moment is aligned with the magnetic field.  In non-relativistic quantum mechanics, this potential energy appears as a term in the Hamiltonian (underlined below) that generates the time evolution of a two-component electron wave function $\chi$ via the Pauli equation,
\begin{align}
i \hbar \frac{\partial \chi}{\partial t} &= \widehat{H}\chi
\nonumber
\\
&= \left[\frac{\left(-i \hbar \vec{\nabla}+\frac{e}{c}\vec{A}\,\right)^2}{2 m}+\underline{\mu_B\: \vec{\sigma}_p \cdot \vec{B}}-e\phi \right]\chi
\ ,
\label{pauli}
\end{align}
where $\phi$ and $\vec{A}$ are the scalar and vector potentials, $\vec{B}=\vec{\nabla}\times\vec{A}$, $\vec{\sigma}_p$ are the Pauli spin matrices
\begin{equation}
\sigma_p^1=\left(\begin{matrix} 0 & 1 \\  1 & 0 \end{matrix}\right)
\quad\quad
\sigma_p^2=\left(\begin{matrix} 0 & -i \\  i & 0 \end{matrix}\right)
\quad\quad
\sigma_p^3=\left(\begin{matrix} 1 & 0 \\ 0 & -1 \end{matrix}\right)
\ ,
\label{matrixdefs}
\end{equation}
and $\mu_B\: \vec{\sigma}_p$ in the underlined term is the operator version of the classical magnetic moment vector $\vec{m}$ (with the strength of the magnetic moment fixed by $\mu_B$).  In the Pauli equation \eqref{pauli}, the fact that the electron has a magnetic moment equal to the Bohr magneton $\mu_B$ appears directly.   In the Dirac equation, by contrast, the magnetic moment cannot be immediately read off and must be derived.

The Dirac equation can be written as
 \begin{equation}
i \hbar \frac{\partial \psi}{\partial t} = \big(-i \hbar c\, \gamma^0\vec{\gamma}\cdot\vec{\nabla} + \gamma^0 m c^2+ e\, \gamma^0\vec{\gamma} \cdot \vec{A}-e\phi \big)\psi
\ .
\label{dirac}
\end{equation}
The gamma matrices that appear in \eqref{dirac} are
\begin{equation}
\vec{\gamma}=\left(\begin{matrix} 0&\vec{\sigma}_p\\ -\vec{\sigma}_p&0 \end{matrix}\right) \quad\quad \gamma^0 = \left(\begin{matrix} I&0\\ 0&-I \end{matrix}\right)
\ ,
\label{matrixdefs2}
\end{equation}
where $I$ is the $2 \times 2$ identity matrix and $\vec{\sigma}_p$ are the $2 \times 2$ Pauli matrices \eqref{matrixdefs}.

The Dirac equation is sometimes interpreted as part of a single-particle relativistic quantum mechanical theory (where it gives the dynamics for a four-component complex-valued electron wave function $\psi$) and sometimes interpreted as part of a relativistic classical field theory (where it gives the dynamics for a four-component complex-valued field $\psi$).  We will be discussing both interpretations in this paper, but for the moment let us follow standard practice for the derivation of the magnetic moment and understand $\psi$ to be a quantum wave function.

We now have the pieces in place to review how the Dirac equation for an electron in an external magnetic field limits to the Pauli equation with the standard Dirac magnetic moment (the Bohr magneton).  In the non-relativistic limit, the $\beta m c^2$ term in \eqref{dirac} leads to a rapid and spatially independent change in the electron wave function's phase that can be factored out by writing the wave function as
\begin{equation}
\psi= \exp\left[-\frac{i}{\hbar} m c^2 t \right] \left(\begin{matrix} \chi_u \\ \chi_l \end{matrix}\right)
\ ,
\label{psidecomposition}
\end{equation}
where $\chi_u$ and $\chi_l$ are two two-component upper and lower pieces that combine to form a four-component entity.  In the end, $\chi_u$ will turn out to be the two-component Pauli electron wave function.

Putting \eqref{dirac}, \eqref{matrixdefs2}, and \eqref{psidecomposition} together yields a dynamic equation for $\chi_u$ and $\chi_l$,
\begin{equation}
i \hbar \frac{\partial}{\partial t} \left(\begin{matrix} \chi_u \\ \chi_l \end{matrix}\right) =-i \hbar c \left(\begin{matrix} \vec{\sigma}_p\cdot\vec{\nabla} \chi_l \\ \vec{\sigma}_p\cdot\vec{\nabla}\chi_u \end{matrix}\right) - 2 m c^2 \left(\begin{matrix} 0 \\ \chi_l \end{matrix}\right) + e\left(\begin{matrix} \vec{\sigma}_p\cdot\vec{A}\: \chi_l \\ \vec{\sigma}_p\cdot\vec{A}\: \chi_u \end{matrix}\right)  - e \phi \left(\begin{matrix} \chi_u \\ \chi_l \end{matrix}\right)
\ .
\label{brokenupDirac}
\end{equation}
If we choose to focus on positive-frequency (positive-energy) solutions to the Dirac equation,\footnote{In the context of relativistic quantum mechanics, the negative-frequency modes are sometimes set aside because they are taken to be filled by an infinite ``Dirac sea'' of negative energy electrons.  Any holes where electrons are missing from the sea would appear as positrons.  For our purposes here, we need not delve into difficult questions about the interpretation of negative frequency modes in relativistic quantum mechanics.  Let us simply note that they can, in one way or another, be reserved for the description of positrons.} then we can assume that $\chi_l$ is varying slowly and express it in terms of $\chi_u$ (using the lower part of \eqref{brokenupDirac} with $\frac{\partial}{\partial t}\chi_l\approx 0$) as
\begin{equation}
\chi_l \approx\left( \frac{- i \hbar c \: \vec{\sigma}_p\cdot\vec{\nabla} + e \: \vec{\sigma}_p\cdot\vec{A}}{2 m c^2+e\phi}\right)\chi_u
\ .
\label{upperlower}
\end{equation}
If we further assume that $e\phi$ is small relative to $m c^2$ (an assumption about the electromagnetic field that can be satisfied by setting $\phi$ to zero for the case at hand of an external magnetic field), then the above expression becomes
\begin{equation}
\chi_l \approx\left( \frac{- i \hbar c \: \vec{\sigma}_p\cdot\vec{\nabla} + e \: \vec{\sigma}_p\cdot\vec{A}}{2 m c^2}\right)\chi_u
\ .
\label{upperlower}
\end{equation}
Looking now at the upper part of \eqref{brokenupDirac} and inserting this expression for $\chi_l$ yields a dynamical equation for $\chi_u$:
\begin{align}
i \hbar \frac{\partial \chi_u}{\partial t} &= \left[\frac{\left(-i \hbar \vec{\nabla}+\frac{e}{c}\vec{A}\,\right)^2}{2 m}+\frac{e\hbar}{2 m c} \vec{\sigma}_p \cdot \vec{B}-e\phi \right]\chi_u
\ ,
\label{pauli2}
\end{align}
where here we have made use of the identity for Pauli matrices,
\begin{equation}
(\vec{\sigma}_p\cdot\vec{a})(\vec{\sigma}_p\cdot\vec{b})=\vec{a}\cdot\vec{b} + i \vec{\sigma}_p\cdot\left(\vec{a}\times\vec{b}\right)
\ ,
\end{equation}
for $\vec{a}=\vec{b}=-i\hbar\vec{\nabla}+\frac{e}{c}\vec{A}$ and the fact that
\begin{align}
\left(-i\hbar\vec{\nabla}+\frac{e}{c}\vec{A}\right)\times\left(-i\hbar\vec{\nabla}+\frac{e}{c}\vec{A}\right)&=\frac{-ie\hbar}{c}\left(\vec{\nabla}\times\vec{A}\right)
\nonumber
\\
&=\frac{-ie\hbar}{c}\vec{B}
\ .
\end{align}
The dynamical equation for $\chi_u$ that we have thus arrived at \eqref{pauli2} can be recognized as our original Pauli equation \eqref{pauli} with the electron's magnetic moment equal to the Bohr magneton \eqref{bohrmagneton} (the Dirac magnetic moment).  This completes the standard derivation of the electron's magnetic moment from the Dirac equation.

The magnetic moment of the electron is sometimes expressed by giving a proportionality constant $g$ relating the spin magnetic moment to the spin angular momentum $S$,
\begin{equation}
\mu=\frac{g e}{2 m c} S
\ .
\label{gfactor}
\end{equation}
If we plug in $S=\frac{\hbar}{2}$ as the spin angular momentum of the electron, the above derivation yields $g=2$ as a characterization of the ratio between the spin magnetic moment and the spin angular momentum (the gyromagnetic ratio).  This is not the true value of $g$, but it is a good approximation.  The calculation of the anomalous magnetic moment in quantum field theory is often presented as a calculation of $g-2$.  To first-order, the calculation yields $g-2=\frac{e^2}{\pi \hbar c}$, as in \eqref{anomalousmagneticmoment}.

\subsection{Method 2: Analyzing the Current Density}\label{M2section}

In classical electromagnetism, the magnetic moment of a charge distribution with current density $\vec{J}$ is\footnote{See \cite[sec.\ 5.6]{jackson}; \cite[sec.\ 11.2]{zangwill2012}.}
\begin{equation}
\vec{m} = \frac{1}{2 c} \int d^3 \vec{x} \left(\vec{x}\times \vec{J}\,\right)
\ .
\label{generalmagneticmoment}
\end{equation}
For example, a sphere of uniform charge density with charge $Q$ and radius $R$ that is rotating about the $z$ axis at a constant angular velocity $\omega$ would have a current density of
\begin{equation}
\vec{J}=\rho \vec{v}=\frac{Q \omega}{\frac{4}{3}\pi R^3} \hat{z} \times \vec{x}
\label{zupcurrentsphere}
\end{equation}
within the bounds of the sphere.  (Here $\hat{z}$ is the unit vector pointing in the $z$ direction.)  Using \eqref{generalmagneticmoment}, the magnetic moment can be calculated to be
\begin{equation}
\vec{m}=\frac{Q \omega R^2}{5 c} \hat{z}
\ .
\label{zupmagneticmoment}
\end{equation}
In this case and in general, the magnetic moment of a rotating body is larger the greater the charge of the body, the faster the rotation, and the farther the charge is located from the axis of rotation---quantities represented respectively by $Q$, $\omega$, and $R$ in \eqref{zupmagneticmoment}.

This method for calculating the magnetic moment can be applied to the electron, provided that we have a current density $\vec{J}$ to analyze.  There is no need to impose an external magnetic field, as in section \ref{M1section}.  For the Dirac equation, the standard charge and current densities are
\begin{align}
\rho &= -e \psi^\dagger \psi
\label{diracchargedensity}
\\
\vec{J} &= -e c \psi^\dagger \gamma^{0} \vec{\gamma} \psi 
\ .
\label{diraccurrentdensity}
\end{align}
If we retain the interpretation of $\psi$ as a quantum wave function from section \ref{M1section}, then the charge and current densities might be taken to capture how the quantum electron contributes to the classical electromagnetic field (by acting as source terms in Maxwell's equations).\footnote{Here we are treating the electron's charge as spread out and flowing, with charge and current densities given by \eqref{diracchargedensity} and \eqref{diraccurrentdensity}.  That is compatible with some approaches to the foundations of quantum physics and incompatible with others \cite{electronchargedensity}.  For example, it is incompatible with a Bohmian theory where the electron is understood to be a point charge that moves within its wave function \cite[sec.\ 6]{howelectronsspin}.  Bohm and Hiley \cite[sec.\ 10.4]{bohmhiley} explore the idea that you might attribute the electron's magnetic moment to the circulating motion of such a point electron (a motion that the Gordon decomposition, presented in this section, plays a role in describing \cite{holland1999, colijn2002}).}  That kind of approach blends quantum physics for the electron with classical physics for the electromagnetic field and thus yields a semiclassical theory.  Another option would be to reinterpret the Dirac equation as part of a classical field theory where $\psi$ is the Dirac field.  The Dirac equation describes how the Dirac field responds to the electromagnetic field, and Maxwell's equations (with the charge and current densities of the Dirac field as source terms) describe how the electromagnetic field responds to the Dirac field.  Then, you would be dealing with a purely classical theory of two interacting fields.  This theory of interacting classical electromagnetic and Dirac fields can be called ``classical Maxwell-Dirac field theory'' or simply ``Maxwell-Dirac theory.''  At a technical level, these two options are the same.\footnote{Still, these two options are distinct and each comes with its own path for extension to quantum field theory \cite[ch.\ 11]{durr2020}; \cite{fields}; \cite[sec.\ 6.5]{tumulka2022}.  One could potentially arrive at a more precise formulation of quantum field theory by picking one of these paths and combining it with an ``interpretation'' of quantum physics (such as the many-worlds interpretation, spontaneous collapse, or Bohmian mechanics).  For example, one might take the path that starts from classical Maxwell-Dirac field theory and pursue a Bohmian quantum field theory where the Dirac and electromagnetic fields are guided by a wave functional.  The analysis of the electron's magnetic moment presented in this paper is compatible with a variety of different strategies for arriving at a more precise formulation of quantum field theory.}  So, for our purposes here, we can keep both interpretive options on the table.

Let us now derive the Gordon decomposition, a useful expansion of the Dirac current density \eqref{diraccurrentdensity}.\footnote{This presentation follows \cite{gordon1928}; \cite[pg.\ 321--322]{frenkel}; \cite[pg.\ 108]{sakuraiAQM}.}  One can derive a version of the Gordon decomposition from the free Dirac equation, but let us include the electromagnetic field and use the full Dirac equation to arrive at the complete Gordon decomposition (as it will be useful later).  From the Dirac equation \eqref{dirac}, it follows that $\psi^\dagger$ evolves by the conjugate equation,
\begin{equation}
i \hbar \frac{\partial \psi^\dagger}{\partial t} = -i \hbar c\big(\vec{\nabla} \psi^\dagger\big)\cdot \gamma^0\vec{\gamma} - m c^2\psi^\dagger \gamma^0- e\, \psi^\dagger \gamma^0\vec{\gamma} \cdot \vec{A}+e\phi\psi^\dagger 
\ .
\label{conjugatediracequation}
\end{equation}
Multiplying by $\gamma^0$ on the right and rearranging yields
 \begin{equation}
\psi^\dagger = \frac{1}{mc^2}\left( - i \hbar \frac{\partial \psi^\dagger}{\partial t}\gamma^0+i \hbar c \big(\vec{\nabla} \psi^{\dagger}\big)\cdot \vec{\gamma} + e\, \psi^\dagger \vec{\gamma}\cdot \vec{A}+e\phi\psi^\dagger \gamma^0\right)
\ .
\label{conjugatediracequationrearranged}
\end{equation}
Similarly, the Dirac equation can be used to express $\psi$ as
 \begin{equation}
 \psi=\frac{1}{mc^2}\left(  i \hbar \gamma^0 \frac{\partial \psi}{\partial t} +i \hbar c\, \vec{\gamma}\cdot\vec{\nabla}\psi- e\, \vec{\gamma}\psi \cdot \vec{A}+e\phi\,\gamma^0\psi \right)
\ .
\label{diracequationrearranged}
\end{equation}
Plugging \eqref{conjugatediracequationrearranged} and \eqref{diracequationrearranged} into the current density \eqref{diraccurrentdensity} yields,
 \begin{align}
\vec{J}&= -e c \psi^\dagger \gamma^{0} \vec{\gamma} \psi
\nonumber
\\
&= -\frac{e c}{2} \psi^\dagger \gamma^{0} \vec{\gamma} \left[\frac{1}{mc^2}\left( i \hbar \gamma^0 \frac{\partial \psi}{\partial t}+i \hbar c\, \vec{\gamma}\cdot\vec{\nabla} \psi - e\, \vec{\gamma}\psi \cdot \vec{A}+e\phi\,\gamma^0\psi \right)\right]
\nonumber
\\
&\quad\quad -\frac{e c}{2}\left[ \frac{1}{mc^2}\left( - i \hbar \frac{\partial \psi^\dagger}{\partial t}\gamma^0+i \hbar c \big(\vec{\nabla} \psi^{\dagger}\big)\cdot \vec{\gamma} + e\, \psi^\dagger \vec{\gamma}\cdot \vec{A}+e\phi\psi^\dagger\gamma^0\right) \right] \gamma^{0} \vec{\gamma} \psi
\nonumber
\\
&=\frac{i e\hbar}{2 m c}\frac{\partial}{\partial t}(\psi^\dagger \vec{\gamma} \psi)-\frac{i e \hbar}{2m} \left[\psi^\dagger \gamma^{0} \vec{\gamma}\big(\vec{\gamma}\cdot\vec{\nabla} \psi\big)+\big(\vec{\nabla} \psi^{\dagger}\big)\cdot \vec{\gamma} \gamma^{0} \vec{\gamma} \psi \right]-\frac{e^2}{m c}\psi^{\dagger}\gamma^0\psi \vec{A}
\ .
\label{expandingcurrent}
\end{align}
The terms involving $\vec{A}$ are combined in the last line above using the anticommutation relations for the gamma matrices,
 \begin{equation}
\{\gamma^{\mu}, \gamma^{\nu}\}=2\eta^{\mu \nu}
\ ,
\label{anticommutation}
\end{equation}
from which it follows that $\gamma^0\gamma^i\gamma^j - \gamma^j\gamma^0\gamma^i= \gamma^0(\gamma^i\gamma^j+\gamma^j\gamma^i)=\gamma^0 (-2\delta^{ij})$.  The next step uses the $4\times 4$ $\vec{\sigma}$ matrices, related to the $2\times 2$ Pauli spin matrices via
\begin{equation}
\vec{\sigma}=\left(\begin{matrix}\vec{\sigma}_p & 0 \\ 0 & \vec{\sigma}_p \end{matrix}\right)\ ,
\label{matrixdefs}
\end{equation}
and the gamma matrices by
\begin{align}
\sigma^{1} &= i \gamma^{2}\gamma^{3}
\nonumber
\\
\sigma^{2}&= i \gamma^{3}\gamma^{1}
\nonumber
\\
\sigma^{3}&= i \gamma^{1}\gamma^{2}
\ .
\end{align}
Using the $\vec{\sigma}$ matrices, the current density in \eqref{expandingcurrent} can be written as
\begin{equation}
\vec{J}= \frac{i e\hbar}{2 m c}\frac{\partial}{\partial t}(\psi^\dagger \vec{\gamma} \psi)+\frac{i e \hbar}{2m} \left[\psi^\dagger \gamma^{0} \vec{\nabla} \psi-\big(\vec{\nabla} \psi^{\dagger}\big)\gamma^{0} \psi \right]- \frac{e\hbar}{2 m} \vec{\nabla}\times(\psi^\dagger \gamma^0 \vec{\sigma}\psi)-\frac{e^2}{m c}\psi^{\dagger}\gamma^0\psi \vec{A}
\ .
\label{gordondecomposition}
\end{equation}
We have thus arrived at the \textbf{Gordon decomposition of the current density}.  Ohanian \cite{ohanian1986} identifies the first term as a polarization current density (associated with spin), the second term as a convection current density (that gives rise to the orbital magnetic moment), and the third term as a magnetization current density (that gives rise to the spin magnetic moment\footnote{See also \cite{howelectronsspin, smallelectronstates}.}).  The fourth term comes from interaction with the vector potential.

To derive the electron's standard spin magnetic moment \eqref{bohrmagneton}, let us analyze the third term, which we can label $\vec{J}_{M}$,
\begin{equation}
\vec{J}_{M}=- \frac{e\hbar}{2 m} \vec{\nabla}\times(\psi^\dagger \gamma^0 \vec{\sigma}\psi)
\ .
\label{spinterm}
\end{equation}
This term takes the form
\begin{equation}
\vec{J}_{M}=c\, \vec{\nabla}\times \vec{M}
\ ,
\label{currenttomagneticmomentdensity}
\end{equation}
where $\vec{M}$ is the magnetic moment density
\begin{equation}
\vec{M}= - \frac{e\hbar}{2 m c} \psi^\dagger \gamma^0 \vec{\sigma}\psi
 \ .
\label{magneticmomentdensity}
\end{equation}
In this magnetic moment density, the prefactor $\frac{e\hbar}{2 m c}$ is the Bohr magneton \eqref{bohrmagneton}.

You can calculate a spin magnetic moment for the electron from the spin magnetic moment term in the current density \eqref{spinterm} equivalently either by integrating \eqref{magneticmomentdensity} over all of space or by using the earlier general expression for the magnetic moment associated with a given current density \eqref{generalmagneticmoment} and plugging in \eqref{spinterm}.\footnote{See \cite[pg.\ 504]{ohanian1986}; \cite[sec.\ 13.2.4]{zangwill2012}.}  Let us take the non-relativistic limit---as in \eqref{psidecomposition}---such that the magnetic moment density \eqref{magneticmomentdensity} becomes
\begin{align}
 \vec{M}&=- \frac{e\hbar}{2 m c} \psi^\dagger \gamma^0 \vec{\sigma}\psi
 \nonumber\\
 &\approx  - \frac{e\hbar}{2 m c} \chi_u^\dagger \vec{\sigma}_p\chi_u 
 \ ,
\end{align}
using \eqref{matrixdefs}.  As a simple illustrative example, consider a $z$-spin up state (where at every location $\sigma^3_p\chi_u= \chi_u$) for which the magnetic moment density simplifies to
\begin{align}
- \frac{e\hbar}{2 m c} \chi_u^\dagger \vec{\sigma}_p\chi_u  &=  - \frac{e\hbar}{2 m c} \chi_u^\dagger \chi_u \hat{z}
\label{NRmagneticmomentdensity}
\end{align}
Because the integral of $\chi_u^\dagger \chi$ over all of space yields 1 for a normalized state,\footnote{On a quantum interpretation of $\psi$, a normalized state is one for which the integral of the probability density $\psi^\dagger \psi$ over all space is 1.  On a classical field interpretation of $\psi$, we might call the state ``normalized'' if the integral of the charge density $-e \psi^\dagger \psi$ is the charge of the electron, $-e$ (a condition that should be met for a state of the field representing a single electron).  At a technical level, these two conditions are the same.\label{normalizationfootnote}} the total spin magnetic moment that one arrives at by integrating the magnetic moment density in \eqref{NRmagneticmomentdensity} is $- \frac{e\hbar}{2 m c} \hat{z}$ (oriented opposite the spin angular momentum).\footnote{For a $z$-spin up state that is very tightly peaked (and thus outside the non-relativistic limit), the magnetic moment that you calculate from the spin magnetic moment term in the current density \eqref{spinterm} can drop below the Bohr magneton \cite{smallelectronstates}.}  This is the standard Dirac magnetic moment: the Bohr magneton \eqref{bohrmagneton}.\footnote{Darwin \cite{darwin1928} gives a similar derivation of the Bohr magneton as the magnetic moment of the electron by analyzing the Dirac current density in the non-relativistic limit without using the Gordon decomposition.}

There is a dependence of the magnetic moment on the state here in the sense that: if one were to consider a state that is not purely $z$-spin up (for which $\sigma^3_p\chi_u$ is not equal to $\chi_u$ at every location), one would arrive at a spin magnetic moment that is different from $- \frac{e\hbar}{2 m c} \hat{z}$.  This kind of state dependence is to be expected, as a superposition of different spins at different locations would not yield the same total spin magnetic moment as a state of consistent definite spin.  In the next section, we will see that once self-interaction and mass renormalization are incorporated, the spin magnetic moment is state-dependent even when we restrict our attention to states that are purely $z$-spin up.

\section{Going Beyond the Dirac Magnetic Moment}\label{BDsection}

We have just seen two ways that the standard Dirac magnetic moment ($g=2$) can be derived from the Dirac equation.  Corrections to this magnetic moment are ordinarily found within quantum field theory, where one can calculate the anomalous magnetic moment ($g-2$) to high precision by incorporating self-interaction and mass renormalization.  However, it is possible to include these two effects without going to quantum field theory.  In this section, we will modify the derivation of the electron's spin magnetic moment from an analysis of the current density in section \ref{M2section} to incorporate these effects.   Then, in section \ref{RM1section} we will return to section \ref{M1section}’s derivation of the magnetic moment from the non-relativistic limit of the Dirac equation.

\subsection{Self-Interaction}\label{SICDsection}

As was discussed earlier, the Dirac equation can be viewed either as an equation governing the dynamics of a quantum wave function interacting with a classical electromagnetic field or the dynamics of a classical Dirac field interacting with a classical electromagnetic field.  Either way, we can incorporate self-interaction by allowing the charge and current densities in \eqref{diracchargedensity} and \eqref{diraccurrentdensity} to act as source terms for the electromagnetic field, resulting in contributions to the vector and scalar potentials of $\vec{A}_{self}$ and $\phi_{self}$ that alter the dynamics via their appearance in the Dirac equation \eqref{dirac}.  Put another way, we can allow different parts of the electron's cloud of charge to interact with one another---with that interaction being mediated by the electromagnetic field.  To fairly compare the predictions of the Dirac equation and quantum field theory (to understand how anomalous the electron's magnetic moment is) we should not neglect self-interaction when studying the Dirac equation.

Let us work in the Coulomb gauge ($\vec{\nabla}\cdot \vec{A}=0$).  In this gauge, the electron's charge density will act as a source for the scalar potential $\phi_{self}$.  The presence of this potential leads to a problematic kind of self-interaction: self-repulsion.  Different parts of the electrons's charge distribution would repel one another.  This is not an effect that we see in nature.  As I understand the situation, that is because self-repulsion is eliminated when one transitions to quantum field theory \cite{selfrepulsion}.  By contrast, the self-interaction with the vector potential $\vec{A}_{self}$ is retained in the move to quantum field theory.  We will focus in this section on that interaction with $\vec{A}_{self}$ and thus on a kind of self-interaction that is preserved in quantum field theory.

The term in the Gordon decomposition \eqref{gordondecomposition} featuring the vector potential $\vec{A}$,
\begin{equation}
-\frac{e^2}{m c}\psi^{\dagger}\gamma^0\psi \vec{A}
\ ,
\label{Aterm}
\end{equation}
is dropped when studying the free Dirac equation (as in \cite[pg.\ 37]{bjorkendrell}; \cite{ohanian1986, howelectronsspin, smallelectronstates}).  But, \eqref{Aterm} is a genuine contribution to the electron's current density that could modify the electron's magnetic moment.  This term describes how the electron's current density is altered by an external electromagnetic field and also how it is altered by the electron's own field.  It yields a self-interaction correction to the current density.  In particular, there will be a contribution to $\vec{A}$ capturing the magnetic field produced by the term in the Gordon decomposition that we earlier associated with the electron's spin magnetic moment \eqref{spinterm}.  This term, describing a flow of electron charge that yields the Dirac magnetic moment, will produce a magnetic field that alters the flow of electron charge in a way that we can calculate to get a better estimate of the electron's magnetic moment.  To see how this effect changes the current density associated with the electron's spin magnetic moment, let us calculate the first-order correction.  (One could later proceed to calculate higher-order corrections by asking how the aforementioned alteration to the current density changes $\vec{A}$, yielding a further adjustment to the current density via \eqref{Aterm} that would then change $\vec{A}$, and so on.  This procedure of finding higher-order corrections parallels the perturbative calculations of the anomalous magnetic moment performed within quantum field theory.)

Let us now calculate how the $\vec{A}$ term in the Gordon decomposition \eqref{Aterm} alters the electron's magnetic moment.  To simplify matters, let us assume that we are dealing with a magnetostatic scenario, where the current density is approximately constant (and we can thus approximate certain interactions that occur over time as instantaneous).  With this assumption in place, the vector potential at a given location can be written in terms of the current density elsewhere (at the same time) as\footnote{See \cite[sec.\ 5.4]{jackson}; \cite[pg.\ 244]{griffiths}.}
\begin{equation}
\vec{A}(\vec{x})=\frac{1}{c}\int d^3 \vec{y} \frac{\vec{J}(\vec{y})}{|\vec{x}-\vec{y}|}
\ .
\label{AfromJ}
\end{equation}
Focusing on the original spin magnetic moment term in the current density $\vec{J}_{M}$ \eqref{spinterm}, we get a contribution to the vector potential of
\begin{equation}
\vec{A}_{1}(\vec{x})=\frac{1}{c}\int d^3 \vec{y} \frac{- \frac{e\hbar}{2 m} \vec{\nabla}_{\vec{y}}\times(\psi^\dagger(\vec{y}) \gamma^0 \vec{\sigma}\psi(\vec{y}))}{|\vec{x}-\vec{y}|}
\ ,
\label{A1fromJ}
\end{equation}
where $\vec{\nabla}_{\vec{y}}\,\times$ is the curl with respect to the $\vec{y}$ coordinates that are being integrated over.  Plugging this into the $\vec{A}$ term in the Gordon decomposition \eqref{Aterm} yields a correction to the current density of
\begin{align}
\vec{J}_{1} &= -\frac{e^2}{m c}\psi^{\dagger}\gamma^0\psi \vec{A}_1
\nonumber
 \\
 &= \frac{e^3 \hbar}{2 m^2 c^2}\int d^3 \vec{y} \frac{\psi^{\dagger}(\vec{x})\gamma^0\psi(\vec{x}) \vec{\nabla}_{\vec{y}}\times(\psi^\dagger(\vec{y}) \gamma^0 \vec{\sigma}\psi(\vec{y}))}{|\vec{x}-\vec{y}|}
 \ ,
 \label{currentdensityshift}
 \end{align}
where the $1$ subscript on $\vec{J}_1$ denotes that this is the first-order correction to the spin magnetic moment current density.  To find the second-order correction, $\vec{J}_1$ could be used to generate a contribution to the vector potential $\vec{A}_1$ via \eqref{AfromJ} and this could be used to derive a second-order self-interaction correction to the current density $\vec{J}_2$ via \eqref{Aterm}, as in \eqref{currentdensityshift}.  The process could be iterated to yield higher order corrections.

If the first-order correction to the current density \eqref{currentdensityshift} had the same form as the original spin magnetic moment current density \eqref{spinterm}, and only differed in its prefactor, then one could simply read off a correction to the original magnetic moment of $\frac{e\hbar}{2 m c}$.  As it is, the current density in \eqref{currentdensityshift} has a complicated form and is not (or, at least, not obviously) expressible as the curl of some magnetic moment density---as in \eqref{magneticmomentdensity}.  Still, one can use the general expression for the magnetic moment generated by a given current density \eqref{generalmagneticmoment} to find a correction to the magnetic moment from $\vec{J}_{1}$.  That correction looks like it will end up being state-dependent (dependent on $\psi$) even for purely $z$-spin up states (going beyond the ordinary kind of state-dependence mentioned at the end of section \ref{M2section}).  We can see this state-dependence explicitly by considering an illustrative example.

Let us analyze a simple Gaussian wave packet $z$-spin up state:\footnote{This state is discussed in \cite{luttinger1948, huang1952}; \cite[pg.\ 39]{bjorkendrell}; \cite{ohanian1986}; \cite{howelectronsspin}.  The state is not ideal for representing a single electron because it is not formed entirely from positive-frequency (electron) modes \cite{howelectronsspin}.  However, it is a simple and convenient state that should suffice for our purposes here.}
\begin{equation}
\psi=\left(\frac{1}{\pi d^2}\right)^{3/4}e^{-|\vec{x}|^2/2d^2}\left(\begin{matrix} 1\\0\\0\\0 \end{matrix}\right)\ ,
\label{ohanianstate}
\end{equation}
where $d$ specifies the width of the wave packet and will be assumed to be much larger than the Compton radius $\frac{\hbar}{m c}$ (a non-relativistic limit\footnote{A very tightly peaked wave function would include significant contributions from high-frequency (high-energy) modes beyond the non-relativistic limit \cite[pg.\ 503]{ohanian1986}; \cite{smallelectronstates}.}).  In free space, this wave packet would expand over time.  For our purposes here, let us assume that (perhaps because of some external potential) the state is (at least approximately) stable and the current density can be treated as constant---so that the scenario is approximately magnetostatic and we can appeal to \eqref{AfromJ}.

\begin{figure}[htb]
\center{\includegraphics[width=12 cm]{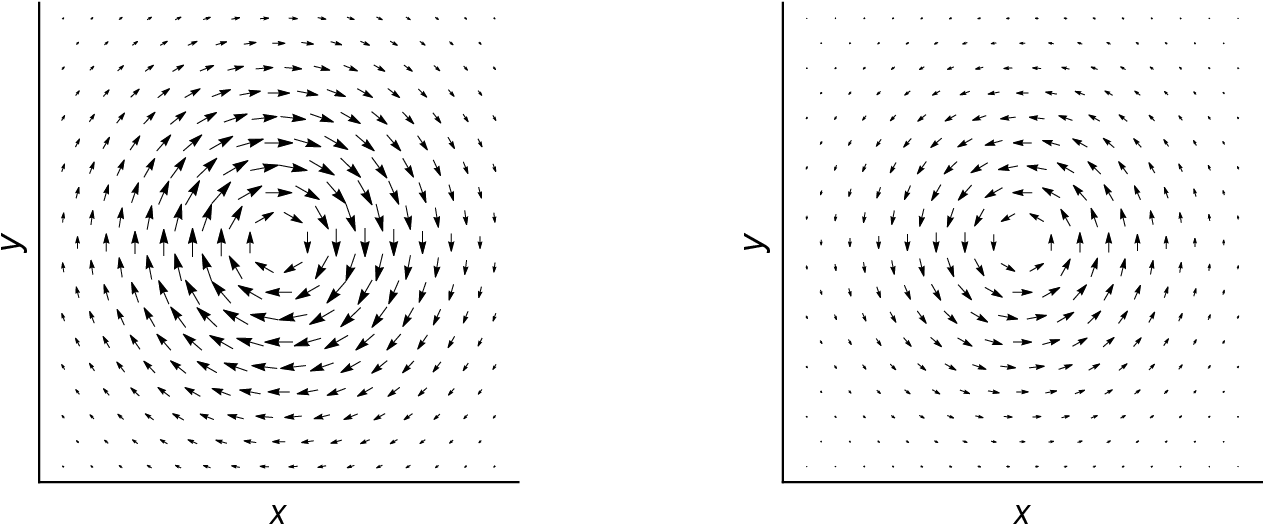}}
\caption{The first plot depicts $\vec{J}_M$, the current density \eqref{magcurrent} associated with the original spin magnetic moment term \eqref{spinterm} for the example $z$-spin up wave packet state \eqref{ohanianstate}, showing only the $xy$ plane at $z=0$.  The second plot (generated by numerical integration) shows the first-order self-interaction correction to the spin magnetic moment current density \eqref{firstordercurrent}, $\vec{J}_1$, pointing opposite $\vec{J}_M$.  The correct size for the arrows in the second plot depends on the wave packet width $d$.  The arrows have been drawn big enough to be easily seen, but they would actually be tiny for a wave packet larger than the Compton radius.}
  \label{currentfigure}
\end{figure}

The original spin magnetic moment term \eqref{spinterm} in the Gordon decomposition of the current density for this state is
\begin{equation}
\vec{J}_M=\frac{e \hbar}{m}\left(\frac{1}{\pi d^2}\right)^{3/2}e^{-|\vec{x}|^2/d^2}\ \frac{\vec{x}\times\hat{z}}{d^2}\ ,
\label{magcurrent}
\end{equation}
describing a flow of charge around the $z$ axis.  This current density alone would yield a total magnetic moment, via \eqref{generalmagneticmoment}, of
 \begin{align}
\vec{m}_0 &= \frac{1}{2 c} \int d^3 \vec{x} \left(\vec{x}\times \vec{J}_M\right)
\nonumber
\\
&= \frac{e \hbar}{2mc}\left(\frac{1}{\pi d^2}\right)^{3/2} \int d^3 \vec{x}\ e^{-|\vec{x}|^2/d^2}\  \frac{(-x_1^2 - x_2^2)}{d^2}\hat{z}
\nonumber
\\
&=  \frac{- e \hbar}{2mc}\hat{z}
\ ,
 \end{align}
with magnitude equal to the Bohr magneton, $|\vec{m}_0|=\frac{e\hbar}{2 m c}$, and orientation opposite the $\hat{z}$ orientation of the spin angular momentum.  Calculating the first-order self-interaction correction to the spin magnetic moment current density, as in \eqref{currentdensityshift}, yields
 \begin{equation}
\vec{J}_1 = -\frac{e^3 \hbar}{m^2 c^2}\left(\frac{1}{\pi d^2}\right)^{3}e^{-|\vec{x}|^2/d^2} \int d^3 \vec{y} \frac{e^{-|\vec{y}|^2/d^2}\ \frac{\vec{y}\times\hat{z}}{d^2}}{|\vec{x}-\vec{y}|}
\ .
\label{firstordercurrent}
 \end{equation}
Numerically integrating to plot this current density at different points gives the appearance of a current density circling the $z$ axis in the opposite direction to $\vec{J}_0$ (as depicted in figure \ref{currentfigure}).

The first-order correction to the current density \eqref{firstordercurrent} gives a first-order correction to the total magnetic moment, via \eqref{generalmagneticmoment}, of
  \begin{align}
\vec{m}_1 &= \frac{1}{2 c} \int d^3 \vec{x}\ \left(\vec{x}\times \vec{J}_1\right)
 \nonumber
 \\
 &= \frac{-e^3 \hbar}{2m^2 c^3} \left(\frac{1}{\pi d^2}\right)^{3} \int d^3 \vec{x} d^3 \vec{y}\ \frac{e^{-\left(|\vec{x}|^2+|\vec{y}|^2\right)/d^2}\ \frac{\vec{x}\times (\vec{y}\times\hat{z})}{d^2}}{|\vec{x}-\vec{y}|}
  \nonumber
 \\
  &=\frac{-e^3 \hbar}{2m^2 c^3} \left(\frac{1}{\pi d^2}\right)^{3} \int d^3 \vec{x} d^3 \vec{y}\ \frac{e^{-\left(|\vec{x}|^2+|\vec{y}|^2\right)/d^2}\ \frac{x_3 \vec{y} - (\vec{x} \cdot \vec{y})\hat{z}}{d^2}}{|\vec{x}-\vec{y}|}
\nonumber
\\
&= \frac{e^3 \hbar}{2m^2 c^3} \left(\frac{1}{\pi d^2}\right)^{3} \int d^3 \vec{x} d^3 \vec{y}\ \frac{e^{-\left(|\vec{x}|^2+|\vec{y}|^2\right)/d^2}\ \frac{(x_1 y_1 + x_2 y_2)}{d^2}\hat{z}}{|\vec{x}-\vec{y}|}
\ .
\label{magneticmomentcorrection}
 \end{align}
To see the dependence on $d$, this integral can be rewritten using the rescaled coordinates $\vec{x}'=\vec{x}/d$ and $\vec{y}\hspace{.5 pt}'=\vec{y}/d$ as
 \begin{align}
\vec{m}_1 &= \frac{e^3 \hbar}{2 m^2 c^3 d} \left(\frac{1}{\pi}\right)^{3} \int d^3 \vec{x}' d^3 \vec{y}\hspace{.5 pt}'\ \frac{e^{-\left(|\vec{x}'|^2+|\vec{y}\hspace{.5 pt}'|^2\right)}\ (x_1' y_1' + x_2' y_2')\hat{z}}{|\vec{x}'-\vec{y}\hspace{.5 pt}'|}
\ .
 \end{align}
From numerical integration, the integral appears to give roughly  $\vec{m}_1 \approx 0.071\frac{e^3 \hbar}{m^2 c^3 d} \hat{z}$.  This first-order self-interaction correction to the magnetic moment scales like $1/d$ and points in the opposite direction to the Dirac magnetic moment $\vec{m}_0$.  The more widely the wave packet is spread out, the smaller the correction.  With the first-order self-interaction correction included, the magnetic moment of the electron is state-dependent (in this case depending on the size of the wave packet).  The first-order correction points in the wrong direction if one hopes to explain the observed anomalous magnetic moment via self-interaction alone (as the observed effect is a slight strengthening of the Dirac magnetic moment).  However, the story does not end here.  In addition to self-interaction, there is another effect that modifies the magnetic moment of the electron in the context of the Dirac equation: mass renormalization.

\subsection{Mass Renormalization}\label{MRsection}

Standard derivations of the electron's anomalous magnetic moment in quantum field theory include both self-interaction and mass-renormalization.  Mass renormalization can be included in the kind of precursor theory that we have been studying, where the electron's dynamics are governed by the Dirac equation.  However, there are questions as to how exactly it should be done.

In such a theory, the electron is surrounded by an electromagnetic field that it itself produces (a self-field).  This self-field will carry a certain amount of energy that, by mass-energy equivalence, contributes to the electron's \emph{dressed} mass $m_e$.  This dressed mass is the sum of the electron's \emph{bare} mass $m_{b}$ and the electromagnetic mass $m_{em}$ of the electron's self-field.\footnote{This kind of mass renormalization (incorporating a mass contribution from the energy of a classical electromagnetic field) is closely related to the mass renormalization that occurs within quantum electrodynamics \cite{weisskopf1939, huang1956}; \cite[pg.\ 242]{huang1998}; \cite{huang2013}; \cite[pg.\ 189, 228--231]{schweber1994}.}

In the Dirac equation, $\psi$ can be regarded either as a quantum wave function or a classical Dirac field.  Either way, let us assume that it is interacting with a classical electromagnetic field as described in section \ref{M2section}.  In this context, we can study the self-field of the electron and calculate the electromagnetic mass\footnote{For more on how the electromagnetic field around a body increases the mass of that body (adding an ``electromagnetic mass'' to the ordinary mass) in the context of classical electromagnetism, see, e.g., \cite[ch.\ 28]{feynman2}; \cite[ch.\ 8]{lange}; \cite{forcesonfields, gravitationalfield}.} by integrating the standard energy density of the electromagnetic field,
\begin{equation}
\rho^{\mathcal{E}}_{em}  = \frac{E^2}{8 \pi}+\frac{B^2}{8 \pi}
\ ,
\label{EMenergydensity}
\end{equation}
and dividing that total electromagnetic energy by $c^2$ to get the electromagnetic mass of the electron.

\begin{SCfigure}
\includegraphics[width=6 cm]{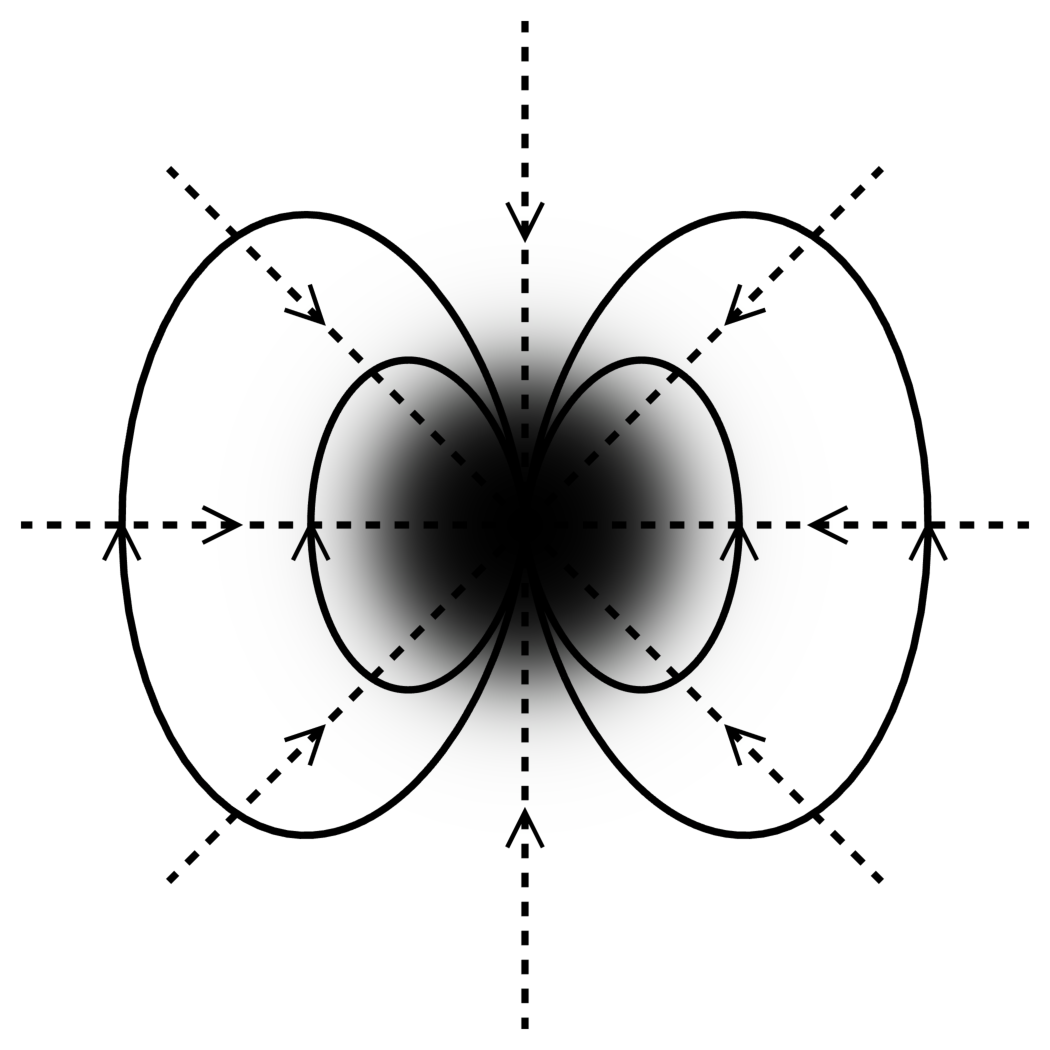}
\caption{This figure gives a rough illustration of a $z$-spin up electron's self-field.  The dark cloud is the electron's charge density.  That charge density rotates about the $z$-axis.  The dotted lines show the Coulomb electric field generated by the charge density, pointing towards the electron (a field that decreases to zero at the center of the electron). The solid lines show the magnetic field generated by the flow of charge, which is similar to that of a small bar magnet (or current loop).}
\label{fieldsfigure}
\end{SCfigure}

In addition to energy density, the electromagnetic field also carries a momentum density that is proportional to $\vec{E} \times \vec{B}$.  Looking at figure \ref{fieldsfigure}, it is clear that for a $z$-spin up electron the momentum density points in circles around the $z$-axis.  This leads to an additional contribution to the angular momentum of the electron that might factor into an accurate calculation of the gyromagnetic ratio---and thus the $g$ factor from \eqref{gfactor}.  We will not explore that complication regarding the spin angular momentum here as the focus of this article is on the electron's spin magnetic moment, not its angular momentum or gyromagnetic ratio.\footnote{For discussion of this issue, see \cite{giulini2008, chalupsky2018, howelectronsspin}.}

Let us return our attention to the self-energy of the electron and the electromagnetic mass that one arrives at by dividing that self-energy by $c^2$.  That electromagnetic mass will depend on the electron's state, which can change over time.  For a $z$-spin up electron wave packet (figure \ref{fieldsfigure}),\footnote{Examining figure \ref{fieldsfigure}, note that the electron's rotating cloud of charge would not yield any electromagnetic radiation, just as a rotating charged sphere does not produce radiation in classical electromagnetism \cite[pg.\ 459]{daboul1973}; \cite[pg.\ 293]{grandy1982}; \cite[sec.\ 6]{chalupsky2018}.  If we think of our context here as a classical Dirac field interacting with a classical electromagnetic field, this yields a classical explanation as to why atoms are stable.  The states of the Dirac and electromagnetic fields that would be superposed to represent the atom in quantum field theory are states where electron charge is flowing but where no radiation would be emitted.  (For more on the spread of electron charge in atoms, see \cite{electronchargedensity}.)} the electric energy density $\frac{E^2}{8 \pi}$ increases as you approach the electron cloud and decreases to zero at the center.  The magnetic energy density $\frac{B^2}{8 \pi}$, by contrast, will be greatest at the center and along the $z$-axis a bit up and down from the center.  The total electromagnetic energy (and thus the electromagnetic mass) will change as the packet evolves.  For example, the electromagnetic mass might shrink as an electron wave packet expands and the self-field around it becomes weaker.  Thus, the procedure of mass renormalization is not as simple as finding a constant mass $m_{em}$ that serves as a consistent difference between $m_e$ and $m_b$.  We will explicitly calculate $m_{em}$ shortly, but before doing so let us continue painting the broad picture as to how mass renormalization might proceed.

To find the bare mass, one could take an approach that parallels what is done for the electromagnetic mass.  There is an energy density\footnote{The idea of assigning an energy density to $\psi$ is sensible if $\psi$ is being interpreted as a classical field.  If $\psi$ is a quantum wave function, it may not be reasonable to assign it an energy density.  But, as we have already assigned $\psi$ a charge density \eqref{diracchargedensity}, we will go ahead and assign it an energy density as well.} for $\psi$ that is like the energy density for the electromagnetic field in \eqref{EMenergydensity},
\begin{equation}
\rho^{\mathcal{E}}_{d}  = \underbrace{\psi^{\dagger}\left(\gamma^0 m c^2\right)\psi}_{\mbox{(a)}} \underbrace{- \frac{i \hbar c}{2}\left(\psi^{\dagger} \gamma^0\vec{\gamma}\cdot\del \psi
 - (\del \psi^{\dagger}) \cdot (\gamma^0\vec{\gamma} \psi )\right)}_{\mbox{(b)}}\underbrace{+ e\, \psi^\dagger\gamma^0\vec{\gamma}\psi \cdot \vec{A}}_{\mbox{(c)}}
\ ,
\label{diracenergydensity}
\end{equation}
where here $m$ is the mass that appears in the Dirac equation (a mass that we shall, at least for the moment, keep distinct from $m_e$, $m_b$, and $m_{em}$).\footnote{This energy density is part of the symmetric energy-momentum tensor for $\psi$.  It appears in \cite[pg.\ 419]{heitler1954}; \cite[pg.\ 156]{hatfield}; \cite[pg.\ 981]{leclerc2006}; \cite[sec.\ 5]{potentialenergy}.}  (The Hamiltonian density of quantum electrodynamics is the sum of operator versions of \eqref{EMenergydensity} and \eqref{diracenergydensity}.\footnote{See \cite[pg.\ 132]{heitler1954}; \cite[pg.\ 87]{bjorkendrellfields}; \cite[sec.\ 8.1]{hatfield}.})  To find the bare mass, one could integrate the energy density \eqref{diracenergydensity} over all of space and divide by $c^2$.  However, arguably terms (b) and (c)\footnote{See \cite[sec.\ 5]{potentialenergy} for a brief discussion as to which terms should be classified as kinetic energy density.} in \eqref{diracenergydensity} should be interpreted as capturing the electron's kinetic energy and thus they should not be taken to contribute to the electron's bare mass $m_b$.  Focusing on (a) and working in the non-relativistic limit, the integral over all space yields $m c^2$ as a total energy and thus $m$ as the bare mass (because in the non-relativistic limit $\psi^{\dagger}\gamma^0\psi \approx \psi^{\dagger}\psi \approx \chi_u^\dagger \chi_u$, and that integrates to 1\footnote{See footnote \ref{normalizationfootnote}.}).  So, if (b) and (c) are kinetic energy density terms, then---in the non-relativistic limit---the bare mass is equal to the mass that appears in the Dirac equation: $m_b=m$.

We have discussed how one might go about calculating the electromagnetic and bare masses $m_{em}$ and $m_b$, and we have seen that (in the non-relativistic limit) $m_{em}$ is state-dependent whereas $m_b$ is not.  For a particular state, we can set the mass $m$ that appears in the Dirac equation to a specific value that is less than the observed electron mass so that the dressed mass, $m_e=m_b+m_{em}$, is equal to the observed mass of the electron (about $9.1\times 10^{-31}$ kilograms).  However, as the state evolves, $m_{em}$ will change and (if $m$ is held fixed) $m_e$ will not remain equal to the observed mass of the electron.  Such are the limits of mass renormalization in this context.\footnote{An alternative option would be to have the mass $m$ that appears in the Dirac equation be time-dependent, chosen at each moment by analyzing the self-field and adjusting to ensure that the sum of the bare mass and the electromagnetic mass is always $m_e$.  This results in a strange theory where the Dirac equation includes a time-dependent mass that depends at every moment on an integral of the electromagnetic self-field over all of space.  That kind of theory will not be explored here.}

If we were to enact mass renormalization at a moment for a particular electron state, we would be revising the mass that appears in the Dirac equation and thus the magnetic moment that we derived from the Dirac equation in section \ref{Dsection}.  Using a Taylor expansion and assuming that $m_{em}$ is small relative to $m_e$, the correction would be
\begin{align}
\mu_B&=\frac{e\hbar}{2 m c}
\nonumber
\\
&=\frac{e\hbar}{2 (m_e-m_{em}) c}
\nonumber
\\
&\approx\frac{e\hbar}{2 m_e c}+\frac{e\hbar m_{em}}{2 m_e^2 c}
\ ,
\label{Taylorexpansion}
\end{align}
to first order in $\frac{m_{em}}{m_e}$.  Because $m_{em}$ is positive, this will lead to an increase in the magnetic moment (that could help explain why the electron's actual magnetic moment is stronger than the Dirac magnetic moment).  To calculate this correction, we must find $m_{em}$.

We can calculate $m_{em}$ by integrating the energy density in \eqref{EMenergydensity}.  This energy density contains contributions from both the electric and magnetic fields.  Let us focus first on the electric field.  If we assume that the electron's charge density \eqref{diracchargedensity} is approximately constant, then we are in an electrostatic scenario\footnote{For the equivalence in \eqref{Eenergydensity}, see, e.g., \cite[sec.\ 8.5]{feynman2}; \cite[sec.\ 3.6]{zangwill2012}} and
\begin{equation}
\int d^3 \vec{x}\ \frac{E^2}{8 \pi} =\frac{1}{2}\int d^3 \vec{x} d^3 \vec{y}\ \frac{\rho(\vec{x})\rho(\vec{y})}{|\vec{x}-\vec{y}|}
\ .
\label{Eenergydensity}
\end{equation}
Let us focus now on the example state from section \ref{SICDsection}, \eqref{ohanianstate}, which---as was discussed earlier---would not be stable in free space but which we will treat as approximately stable for our purposes here.  Inserting the charge density \eqref{diracchargedensity} for this state into \eqref{Eenergydensity} yields
\begin{equation}
\frac{e^2}{2}\left(\frac{1}{\pi d^2}\right)^{3}\int d^3 \vec{x} d^3 \vec{y}\ \frac{e^{-|\vec{x}|^2/d^2}e^{-|\vec{y}|^2/d^2}}{|\vec{x}-\vec{y}|}
\ .
\label{Eenergy}
\end{equation}
This total energy in the electric field will depend on the spread of the state.  Numerical integration suggests the result is something like $0.403\frac{e^2}{d}$, keeping three figures after the decimal place as was done below \eqref{magneticmomentcorrection}.  To understand the inverse dependence on $d$, imagine decreasing $d$.  As the cloud of electron charge shrinks, the electric field near the cloud becomes stronger and the energy stored within that field rises \eqref{Eenergydensity}.

Next, let us calculate the contribution to the electromagnetic energy from the magnetic field.  If we assume that the electron's current density \eqref{diraccurrentdensity} is approximately constant, then we are in a magnetostatic scenario and\footnote{See \cite[pg.\ 214]{jackson}; \cite[eq.\ 26]{potentialenergy}.}
\begin{equation}
\int d^3 \vec{x}\ \frac{B^2}{8 \pi} =\frac{1}{2c}\int d^3 \vec{x}\ \vec{J}\cdot\vec{A}
\ .
\label{Benergydensity}
\end{equation}
Plugging in the spin term in the Gordon decomposition of the current density \eqref{magcurrent} and the associated vector potential \eqref{A1fromJ} for the example state \eqref{ohanianstate} gives,
\begin{align}
\frac{1}{2c}\int d^3 \vec{x}\ \vec{J}_M\cdot\vec{A}_1&= \int d^3 \vec{x} \ \frac{1}{2 c} \left[\frac{e \hbar}{m}\left(\frac{1}{\pi d^2}\right)^{3/2}e^{-|\vec{x}|^2/d^2}\ \frac{\vec{x}\times\hat{z}}{d^2}\right]\cdot\left[\frac{1}{c}\int d^3 \vec{y} \frac{\frac{e \hbar}{m}\left(\frac{1}{\pi d^2}\right)^{3/2}e^{-|\vec{y}|^2/d^2}\ \frac{\vec{y}\times\hat{z}}{d^2}}{|\vec{x}-\vec{y}|}\right]
\nonumber
\\
&=\int d^3 \vec{x} d^3 \vec{y}\ \left(\frac{e^2 \hbar^2}{4 m^2 c^2 d^2} \right)\left(\frac{1}{\pi d^2}\right)^3\frac{e^{-\left(|\vec{x}|^2+|\vec{y}|^2\right)/d^2}\ \frac{(x_1 y_1 + x_2 y_2)}{d^2}}{|\vec{x}-\vec{y}|}
\ ,
\label{Benergy}
\end{align}
Making use of the fact that $(\vec{x}\times\hat{z})\cdot(\vec{y}\times\hat{z})=(\vec{x}\cdot\vec{y}) - (\vec{x}\cdot\hat{z})(\vec{y}\cdot\hat{z})=x_1 y_1 + x_2 y_2$.  The form of this integral is exactly as in \eqref{magneticmomentcorrection} and we can reuse the same numerical integration, yielding approximately $0.036 \frac{e^2 \hbar^2}{m^2 c^2 d^3}$.  Like the electric contribution, the magnetic energy decreases as the size of the wave packet $d$ increases.  However, the decrease is much more rapid.  If we assume (as is appropriate to the non-relativistic limit) that $d$ is large relative to the Compton radius, $\frac{\hbar}{m_e c}$, then for the purposes of our first-order estimate as to how mass renormalization alters the magnetic moment, we can ignore the magnetic self-energy.

Dividing the electric self-energy---presented under \eqref{Eenergy}---for the sample state \eqref{ohanianstate} by $c^2$ gives a rough estimate of the electromagnetic mass:
\begin{equation}
m_{em} \approx 0.403\frac{e^2}{c^2 d}
\ .
\label{EMmassestimate}
\end{equation}
For $d \geq \frac{\hbar}{m_e c}$, this yields $m_{em} \leq 0.403 \frac{e^2 m_e}{\hbar c} \approx 0.0004 m_e$.  That electromagnetic mass is small compared to $m_e$, as was assumed when performing the Taylor expansion in \eqref{Taylorexpansion}.  Applying that Taylor expansion, we get a first-order correction to the magnetic moment $\vec{m}$ of
\begin{equation}
\vec{m}_{em1}= \frac{e\hbar m_{em}}{2 m_e^2 c} \approx 0.403\frac{e^3 \hbar}{2 m_e^2 c^3 d}
\ .
\label{EMcorrection}
\end{equation}
This can then be combined with the Dirac magnetic moment and the self-interaction correction from section \ref{SICDsection}, under \eqref{magneticmomentcorrection}, to get a total magnetic moment (to first-order) within the context of the Dirac equation of
\begin{align}
\vec{m} &= \left[\frac{e\hbar}{2 m_e c}  + (0.403-0.071) \frac{e^3 \hbar}{2 m_e^2 c^3 d} \right] \hat{z}
\nonumber
\\
&= \frac{e\hbar}{2 m_e c}\left[1  + 0.332 \frac{e^2}{m_e c^2 d} \right] \hat{z}
\ .
\label{totalnewmoment}
\end{align}

The key takeaway here is that the magnetic moment is not simply the Dirac magnetic moment, there are state-dependent corrections from self-interaction in the current density and mass renormalization.  For the right choice of wave packet width, $d \approx 2.09 \frac{\hbar}{m_e c}$ (just over twice the Compton radius), \eqref{totalnewmoment} agrees with the first-order prediction of quantum field theory \eqref{anomalousmagneticmoment}.  However, we cannot simply put that wave packet width in by hand.  In this theory, the wave packet representing an electron will evolve over time and not always remain in the same state.

\section{Returning to the Non-Relativistic Limit of the Dirac Equation}\label{RM1section}

We have just seen how the derivation of the Dirac magnetic moment from an analysis of the current density in section \ref{M2section} can be modified to incorporate self-interaction and mass renormalization, yielding a state-dependent magnetic moment \eqref{totalnewmoment}.  One could skip ahead to the comparison with other authors who take similar approaches in section \ref{comparisonsection}, but for those who are interested in tying up loose ends, let us take a moment to see that the derivation of the Dirac magnetic moment from the non-relativistic limit of the Dirac equation in section \ref{M1section} can similarly be modified to incorporate self-interaction and mass renormalization.  The procedure for mass renormalization would be exactly as in section \ref{MRsection}.  What about self-interaction?

Let us return to the Hamiltonian in the non-relativistic limit \eqref{pauli2} acting on the two-component electron wave function $\chi_u$.\footnote{To derive that equation, \eqref{pauli2}, we assumed that $e\phi$ is small relative to $m c^2$.  This assumption was easily satisfied when we were ignoring the self-field and considering only an external electromagnetic field.  Including the self-field, that assumption might be violated.  You could have a situation where $e\phi_{self}$ is not small relative to $m c^2$.  Working in the Coulomb gauge, this might happen for a very tightly peaked charge distribution.  Let us assume that the electron charge distribution is sufficiently widely spread out that $e\phi_{self}$ is everywhere small relative to $m c^2$ (as part of our assumption that we are working in the non-relativistic limit).  If that is so, we can continue to use \eqref{pauli2}.}  There is an $|\vec{A}|^2$ term in that Hamiltonian, $\frac{e^2}{2 m c^2}|\vec{A}|^2$.  To include self-interaction, we must break the vector potential $\vec{A}$ into a self-field piece, $\vec{A}_{self}$, and an external-field piece, $\vec{A}_{ext}$.  The cross term
\begin{equation}
\frac{e^2}{m c^2} \vec{A}_{ext}\cdot\vec{A}_{self}
\label{crossterm}
\end{equation}
in 
\begin{align}
\frac{e^2}{2 m c^2}|\vec{A}|^2 &=\frac{e^2}{2 m c^2}\left(|\vec{A}_{ext}|^2+2 \vec{A}_{ext}\cdot\vec{A}_{self} +|\vec{A}_{self}|^2\right)
\label{wholething}
\end{align}
introduces a new coupling that could yield a correction to the coupling in the $\frac{e\hbar}{2 m c} \vec{\sigma}_p \cdot \vec{B}_{ext}$ term in \eqref{pauli2} (that we had earlier taken to reveal the electron's spin magnetic moment).\footnote{If we were interested in calculating higher-order self-interaction corrections to the spin magnetic moment, we should also look at the $|\vec{A}_{self}|^2$ term in \eqref{wholething} because the self-field vector potential is altered by self-interaction, as was discussed in section \ref{SICDsection}.}  If we assume that the electron is in a uniform external magnetic field $\vec{B}_{ext}$, the external vector potential can be expressed as $\vec{A}_{ext}=\frac{1}{2}\vec{B}_{ext}\times\vec{x}$.\footnote{This way of expressing the vector potential appears in, e.g., \cite[pg.\  4407]{barutdowling1988}; \cite[pg.\ 125]{greiner2000}.} If we assume that the electron current density has been approximately constant, then we can use \eqref{AfromJ} to rewrite $\vec{A}_{self}$ in terms of $\vec{J}$, and the cross term \eqref{crossterm} becomes
\begin{align}
\frac{e^2}{m c^2} \vec{A}_{ext}\cdot\vec{A}_{self}&=\frac{e^2}{2 m c^3}\left(\vec{B}_{ext}\times\vec{x}\right) \cdot \left(\int d^3 \vec{y}\ \frac{\vec{J}(\vec{y})}{|\vec{x}-\vec{y}|}\right)
\nonumber
\\
&=\frac{e^2}{2 m c^3} \vec{B}_{ext} \cdot \left(\int d^3 \vec{y}\ \frac{\vec{x}\times\vec{J}(\vec{y})}{|\vec{x}-\vec{y}|}\right)
\ ,
\label{crossterm2}
\end{align}
making use of the triple product identity $(\vec{a}\times\vec{b}\,)\cdot\vec{c}=\vec{a}\cdot(\vec{b}\times\vec{c}\,)$.  This coupling takes a broadly similar form to the coupling in \eqref{pauli2} associated with the spin magnetic moment, $\frac{e\hbar}{2 m c} \vec{\sigma}_p \cdot \vec{B}_{ext}$, though it does not simply correct the prefactor that was earlier identified as the spin magnetic moment.

To see how this term changes the magnetic moment, we can calculate the expectation value of the two terms together:
\begin{align}
\left\langle \frac{e\hbar}{2 m c} \vec{\sigma}_p \cdot \vec{B}_{ext} + \frac{e^2}{m c^2} \vec{A}_{ext}\cdot\vec{A}_{self} \right\rangle
\ .
\label{totalexpectationvalue}
\end{align}
The first piece is
\begin{align}
\left\langle \frac{e\hbar}{2 m c} \vec{\sigma}_p \cdot \vec{B}_{ext} \right\rangle &= \frac{e\hbar}{2 m c} \int d^3 \vec{x}\ \chi_u^\dagger(x)\vec{\sigma}_p\chi_u(x)\cdot \vec{B}_{ext}
\ ,
\end{align}
which for a $z$-spin up state simplifies to
\begin{align}
\left\langle \frac{e\hbar}{2 m c} \vec{\sigma}_p \cdot \vec{B}_{ext} \right\rangle &= \frac{e\hbar}{2 m c} \hat{z} \cdot \vec{B}_{ext}
\ .
\end{align}
Comparing that expression to \eqref{EMpotentialenergy}, we can read off a Dirac magnetic moment of
\begin{equation}
\vec{m}_0= -\frac{e\hbar}{2 m c} \hat{z}
\ .
\end{equation}

The self-interaction correction to the Dirac magnetic moment comes from the second piece of \eqref{totalexpectationvalue},
\begin{align}
\left\langle \frac{e^2}{m c^2} \vec{A}_{ext}\cdot\vec{A}_{self} \right\rangle&=\frac{e^2}{2 m c^3} \vec{B}_{ext} \cdot \left(\int d^3 \vec{x} d^3 \vec{y}\ \chi_u^\dagger(x)\chi_u(x)\frac{\vec{x}\times\vec{J}(\vec{y})}{|\vec{x}-\vec{y}|}\right)
\ .
\end{align}
Comparing again to \eqref{EMpotentialenergy}, the correction to the magnetic moment is
\begin{equation}
-\frac{e^2}{2 m c^3}\int d^3 \vec{x} d^3 \vec{y}\ \chi_u^\dagger(x)\chi_u(x)\frac{\vec{x}\times\vec{J}(\vec{y})}{|\vec{x}-\vec{y}|}
\ .
\end{equation}
To find the first-order correction, let us take plug in for $\vec{J}$ only the spin term $\vec{J}_M$ \eqref{spinterm} from the Gordon decomposition of the current density.  For the simple sample state studied in section \ref{BDsection}, \eqref{magcurrent}, this yields a first-order correction of
  \begin{align}
\vec{m}_1 &= \frac{-e^3 \hbar}{2m^2 c^3} \left(\frac{1}{\pi d^2}\right)^{{3}} \int d^3 \vec{x} d^3 \vec{y} \frac{e^{-\left(|\vec{x}|^2+|\vec{y}|^2\right)/d^2}\ \frac{\vec{x}\times (\vec{y}\times\hat{z})}{d^2}}{|\vec{x}-\vec{y}|}
\ .
 \end{align}
This is exactly the same first-order self-interaction correction to the spin magnetic moment as was calculated in section \ref{SICDsection}, \eqref{magneticmomentcorrection}.  Thus, at least to first-order, we can incorporate self-interaction and mass renormalization corrections to either of the derivations the Dirac magnetic moment from section \ref{Dsection} and we get the same result.

\section{Comparison to Similar Approaches}\label{comparisonsection}

Other authors have sought to better understand the electron's anomalous magnetic moment by studying the electron's behavior in theories that are simpler than quantum field theory.

Grandy and Aghazadeh \cite[sec.\ 3]{grandy1982} analyze a classical model of the electron as a spherical shell of charge rotating at constant angular velocity.\footnote{Giulini \cite[sec.\ 3.3]{giulini2008} also considers a classical spherical shell of charge rotating at a constant angular velocity and possible values of the gyromagnetic ratio ($g- 2$).  However, Giulini focuses on the angular momentum of the self-field and not on ways in which self-interaction or mass renormalization might alter the magnetic moment.  Chalupsky \cite{chalupsky2018} similarly analyzes the angular momentum of the self-field for a classical rotating Gaussian charge distribution to see how it alters the $g$-factor.}  They place this charge in an external magnetic field and examine the way that radiation reaction alters the Larmor precession, yielding a decrease in the magnetic moment from self-interaction (just as we saw a decrease from self-interaction in section \ref{SICDsection}). Citing Grotch and Kazes \cite{grotch1977} for the importance of including mass renormalization when calculating the anomalous magnetic moment,\footnote{Grotch and Kazes \cite{grotch1977} work in the context of a non-relativistic quantum treatment of the electron interacting with a quantized electromagnetic field, deriving a cutoff-dependent anomalous magnetic moment for the electron with a negative contribution ``due to virtual photon emission and absorption'' (pg.\ 621) and a positive contribution from mass renormalization (just as we found a negative contribution in section \ref{SICDsection} and a positive contribution in section \ref{MRsection}).  As will be discussed shortly, Barut \emph{et al.}\ \cite{barutdowling1988} include mass renormalization (following \cite{grotch1977}) and also derive a cut-off dependent anomalous magnetic moment.} Grandy and Aghazadeh next incorporate mass renormalization (as in section \ref{MRsection}) which yields an increase in the magnetic moment.  Putting these effects together, Grandy and Aghazadeh derive a radius-dependent anomalous magnetic moment for their classical electron that agrees with the standard first-order anomalous magnetic moment \eqref{anomalousmagneticmoment} if you set the radius to one-third of the Compton radius, $\frac{\hbar}{3 m_e c}$ (just as \eqref{totalnewmoment} could be tuned to yield the correct anomalous magnetic moment for the right choice of $d$).  The general conclusions that Grandy and Aghazadeh reach are in line with those reached here.  However, the spherical shell of charge model of the electron does not fit well with quantum field theory.  By contrast, there are clear paths from the Dirac equation (with self-interaction) to quantum field theory.

In a series of paper, Barut and his collaborators have explored a theory that they call ``self-field quantum electrodynamics,'' where you use the Dirac equation to describe the behavior of a single electron and allow the electron to interact with its own electromagnetic self-field \cite{barut1988, barut1989, barut1991}.  This is often presented as a competitor to ordinary quantum electrodynamics, but it is largely in line with the theoretical context described in section \ref{M2section} and can alternatively be seen as a precursor to ordinary quantum electrodynamics (as either a semiclassical theory where a quantum electron wave function interacts with a classical electromagnetic field, or, a classical field theory where the Dirac and electromagnetic fields interact).  Barut and his collaborators suggest eliminating the electromagnetic field from the fundamental ontology.  They argue that if the electromagnetic field is not endowed with any independent degrees of freedom, it can be uniquely reconstructed from the behavior of $\psi$ and thus does not need to be included in the ontology.  Excising the electromagnetic field yields a theory where $\psi$ interacts only with itself.  Taking an alternative approach, one might attempt to eliminate $\psi$ and keep only the electromagnetic field \cite{akhmeteli2022}.  For our purposes here, let us not eliminate either.

Within self-field quantum electrodynamics, Barut \emph{et al.}\ \cite{barutdowling1988} present a non-relativistic derivation of the electron's anomalous magnetic moment and Barut and Dowling \cite{barutdowling1989} present a relativistic derivation.  These papers are insightful and have informed the approach presented here, but they use different methods, make different approximations, and arrive at different conclusions.  Let us consider the two papers in turn.

As in sections \ref{BDsection} and \ref{RM1section}, Barut \emph{et al.}\ \cite{barutdowling1988} work in the Coulomb gauge and calculate corrections to the magnetic moment from both self-interaction and mass renormalization, finding that the self-interaction correction leads to a decrease in the magnetic moment and the correction from mass renormalization leads to a (larger) increase in the magnetic moment.  However, the corrections that they arrive at are dependent on a cutoff removing high-frequency modes of the electromagnetic field and are not dependent on the electron state.  Their analysis is complicated and I will not review it here.  Let me just offer few comparative remarks:  First, Barut \emph{et al.}\ focus on analyzing pieces of the total action density, whereas we have not needed to introduce the action at all here.  Second, they work with an external magnetic field (as in section \ref{RM1section} but unlike section \ref{BDsection}).  Third, as in sections \ref{M2section} and \ref{BDsection}, Barut \emph{et al.}\  \cite[eq.\ 10 and 17]{barutdowling1988} write out the Gordon decomposition of the current density (in a non-relativistic form that can be applied to the current density of the two-component Pauli wave function, or field, $\chi$, from section \ref{M1section}) and focus on the same piece of the Gordon decomposition \eqref{Aterm} when analyzing their correction from self-interaction.  Fourth, in their calculation of the self-interaction correction, Barut \emph{et al.}\ \cite[eq.\ 24]{barutdowling1988} focus on the $\vec{\sigma}_p \cdot \vec{B}_{self}$ term in \eqref{pauli2} instead of the $\vec{A}_{ext}\cdot\vec{A}_{self}$ term (as in section \ref{RM1section}).  Fifth, Barut \emph{et al.}\ calculate the mass renormalization correction by analyzing the $\vec{A}_{self} \cdot \vec{\nabla}$ term in \eqref{pauli2} instead of by integrating the energy density of $\psi$ over all of space (as in section \ref{MRsection}).  Sixth, Barut \emph{et al.}\ \cite[eq.\ 23]{barutdowling1988} assume that the electron is (at least to a first-order of approximation) in a particular state and they do not allow that state to vary (thus missing the state-dependence of the anomalous magnetic moment in this context).  Seventh, that particular state is difficult to interpret as a possible electron state and appears to not be normalizable \cite{bb1986, barut1986}; \cite[pg.\ 156]{grandy1991}; \cite[pg.\ 234]{grandy1991b}.

Barut and Dowling \cite{barutdowling1989} move beyond the non-relativistic limit and argue that the electron's anomalous magnetic moment, at least to first-order, can be correctly calculated within relativistic self-field quantum electrodynamics.  That would be exciting if it were true, but there are reasons to be skeptical.  A couple remarks:  First, Barut and Dowling's use of what they call ``the usual Stueckelberg-Feynman Green's function" \cite[pg.\ 1053]{barutdowling1989} leads to a complex-valued electromagnetic field \cite{bb1986, grandy1991}.  This does not match standard classical electromagnetic field theory (or standard Maxwell-Dirac theory), but Barut \cite[pg.\ 3503]{barut1986} argues that a complex electromagnetic field is appropriate within his self-field quantum electrodynamics.  Second, Grandy \cite[appendix]{grandy1991} has raised a number of technical problems with Barut and Dowling's \cite{barutdowling1989} derivation of the electron's anomalous magnetic moment (to first-order).

\section{Conclusion}

The electron's spin magnetic moment can be calculated to astonishing accuracy within quantum field theory.  That is a massive improvement on any simpler theory that might be viewed as a precursor to quantum field theory.  In section \ref{Dsection}, we saw two ways to use the Dirac equation to arrive at the prediction that the electron's magnetic moment should be the Bohr magneton, a prediction that is close but inaccurate.  Incorporating self-interaction and mass renormalization yields a precursor theory that is closer to quantum field theory---a theory that might be regarded as a semi-classical quantum theory or a classical theory of two interacting fields.  In sections \ref{BDsection} and \ref{RM1section}, we saw that in such a theory the electron's magnetic moment is state-dependent (even when we restrict our attention to, say, purely $z$-spin up states).  Thus, we have arrived at a new puzzle for future research: How does the move from such a precursor to quantum field theory take you from a state-dependent magnetic moment to a fixed magnetic moment?  Put another way:  How does quantum field theory nail down the electron's magnetic moment?

If we can understand how that nailing down is done, it could help us to better understand the anomalous magnetic moment in quantum electrodynamics.  To see how, let us return to the four reasons why one might be dissatisfied by standard derivations of the anomalous magnetic moment within quantum field theory that were identified in the introduction.  The analysis of the Dirac equation presented in section \ref{Dsection} improves on all four points.  First, by studying the electron's current density we can determine its anomalous magnetic moment without placing it in an external magnetic field.  Second, although in this context there are interactions that occur over time and the electron's current density can change, the electron's anomalous magnetic moment is a feature that it possesses in virtue of its state at a single moment---calculated via \eqref{generalmagneticmoment}.  Third, the electron's charge is spread out, avoiding myriad problems facing point charges.  This extended electron fits well with a field wave functional approach to quantum field theory and differs markedly from the point electron depicted in familiar Feynman diagrams (like those in figure \ref{feynmanfigure}).  Fourth, the analysis in section \ref{Dsection} yields a clear physical picture as to why the electron has an anomalous magnetic moment: (i) the electron feels the electromagnetic field that it's ordinary (non-amalous) magnetic moment produces and this self-interaction alters its current density (figure \ref{currentfigure}), and (ii) some of the electron's mass is in the electromagnetic field that surrounds it and thus the constant $m$ that appears in the Dirac equation is not the observed mass of the electron.  These four improvements could potentially be carried over into quantum field theory, but to do so we must work to understand how calculations of the anomalous magnetic moment that can be performed within quantum field theory relate to the calculations that can be done in the context of the Dirac equation.

\vspace*{12 pt}
\noindent
\textbf{Acknowledgments}
Thank you to Andy Akhmeteli, Jacob Barandes, Alexander Blum, Francisco Calder\'{o}n, Adam Koberinski, Logan McCarty, and the anonymous reviewers for helpful feedback and discussion.

\end{document}